\documentstyle[epsfig,float]{elsart}

\begin{document}               

\begin{frontmatter}

\title{Finesse and mirror speed measurement for a suspended
Fabry-Perot cavity using the ringing effect}

\author{Luca Matone\thanksref{ad}},
\author{Matteo Barsuglia},
\author{Fran\c{c}ois Bondu\thanksref{nice}},
\author{Fabien Cavalier},
\author{Henrich Heitmann\thanksref{nice}},
\author{Nary Man\thanksref{nice}}

\address{Laboratoire de l'Acc\'el\'erateur Lin\'eaire,\\
Universit\'e Paris-Sud, Bat 208, 91405 Orsay (France)}

\thanks[ad]{Corresponding author. Present address: 
MS 18-34 LIGO, Caltech, Pasadena CA 91125 (USA); tel.(626) 395-2071;
fax (626) 304-9834; lmatone@ligo.caltech.edu}
\thanks[nice]{Pre.address: Observatoire C\^ote d'Azur,
B.P. 4229, 06304 Nice Cedex 4 (France)}


\begin{abstract}

We here present an investigation of the ringing effect observed on the
VIRGO mode-cleaner prototype MC30. The results of a numerical calculation
show how a simple empirical formula can determine the cavity
expansion rate from the oscillatory behavior. We also show how
the simulation output can be adjusted to estimate the finesse
value of the suspended cavity.
\end{abstract}

\end{frontmatter}

\section{Introduction}

Interferometric gravitational wave detectors, like VIRGO~\cite{virgo}, 
LIGO~\cite{ligo}, GEO~\cite{geo} and TAMA~\cite{tama}, make use of
suspended Fabry-Perot cavities for their properties: spatial filtering
of the laser beam, optical path amplification and power recycling to
name a few. Due to the suspension system, the motion of 
each mirror is dominated by oscillations at the pendulum's fundamental
resonance, typically below $1\,Hz$ and with amplitudes as large as tens
of a laser wavelength.

A VIRGO mode-cleaner prototype MC30 \cite{mc}, which operated in Orsay
for several years, consists of a $30\,m$ triangular Fabry-Perot 
cavity with two possible finesse values, 100 or 1600, depending on
the incoming laser beam polarization state. The two-stage suspension system,
together with a local damping control system, results in a residual
RMS displacement value, for each mirror, of $0.8\,\mu m$.

\begin{figure}
\centering
\epsfig{file=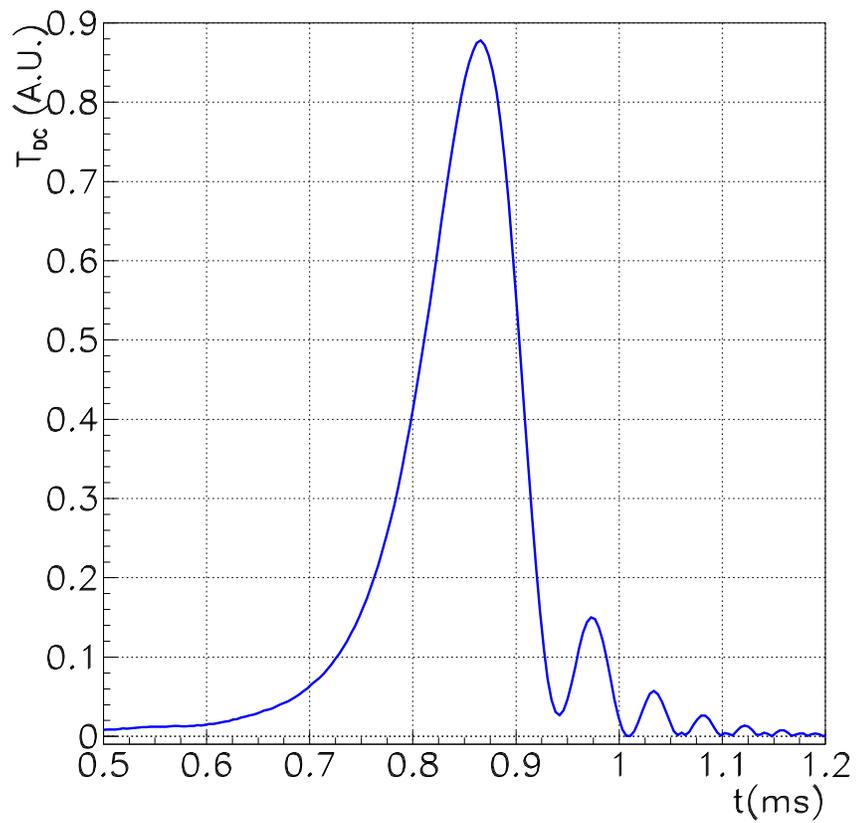,width=12cm}
\caption{Matone et al., Phys.Lett.A} \label{fig:cl9}
\end{figure}

Prior to the lock acquisition, the cavity length sweeps the
optical resonance at different rates of expansion. If the relative
velocity between the mirrors is constant, the DC transmitted power
delineates the {\it Airy peak} as a function of time, easily
observed for the optical system with ${\cal F} = 100$. However,
in the case of ${\cal F} = 1600$, a deformation of the Airy peak,
similar to a ringing, was observed (see fig.(\ref{fig:cl9})).

Both \cite{lefloch,stephane}, and references therein, discuss this
phenomenon. Briefly, this effect arises once the cavity sweeps the
optical resonance in a time $\tau_{sw}$ of the order of
or less than the cavity storage time  $\tau_{st} = 2\,L_0\,{\cal F}
/ c\,\pi$, where $L_0 = 30\,m$ is the cavity length and $c$ the speed of light.

This effect is observed when the rate of expansion is so high
that, as resonance is approached, the cavity doesn't have enough
time to completely fill itself. It is the beating between the
incoming laser field and the evolving stored field that gives rise
to this oscillatory behavior.

The goal of this letter is to present the informations which can be extracted 
from such deformation, in particular, the finesse of the cavity
and the relative mirror speed.


\section{The ringing effect in Fabry-Perot cavities}
 
\begin{figure}
\centering
\epsfig{file=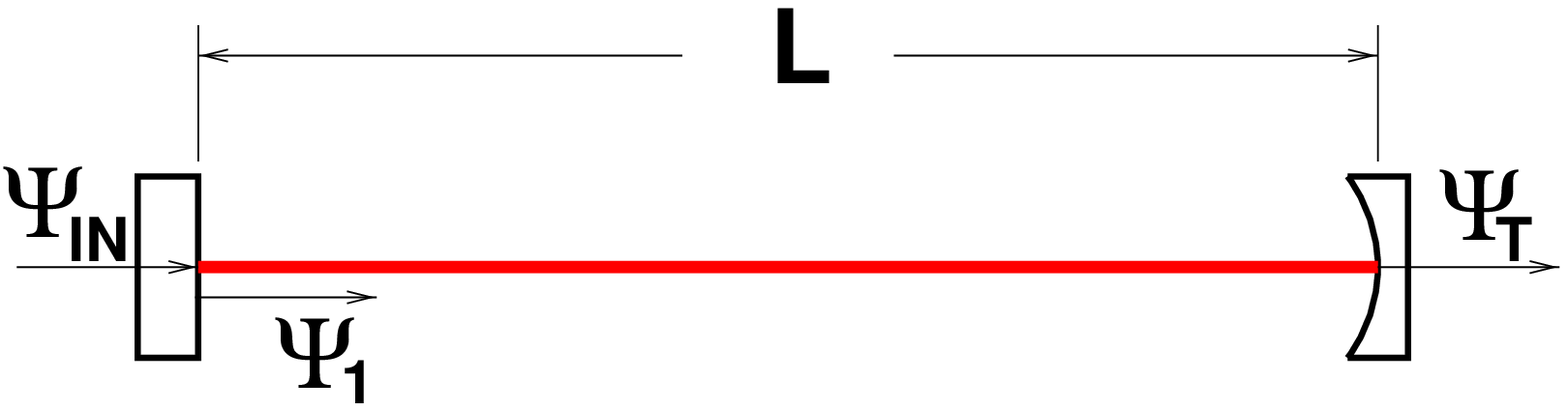,width=12cm}
\caption{Matone et al., Phys.Lett.A}
\label{fig:mc_tran}
\vskip 10cm
\end{figure}

The model used for this study is shown in fig.(\ref{fig:mc_tran}).
Assuming a negligible mirror displacement for times of the order
of the round trip time of light $\tau = 2 L_0 / c = 0.2  \mu s$,
the stored field $\Psi_1(t)$ at time $t$ can be written as
\begin{equation}
\label{eqn:psi1_dyn}
\Psi_1(t)\, =\, t_1\, \Psi_{in}\, +\,
 r_1^{ 2}\, \exp( -\, 2\, i\, k\, L\, )\;\Psi_1(t\, -\, \tau)
\end{equation}
where $r_1$ and $t_1$ denote the amplitude reflectivity and
transmittivity of each mirror,
$\Psi_{in}$ is the incoming laser field, and $L$ is the cavity
length. Assuming that the cavity expands at a constant rate $v$,
we can write $L  = L_0 +  v\, t $
and solve eq.(\ref{eqn:psi1_dyn}) iteratively, for different
velocities $v$ and finesse ${\cal F}$.

Fig.(\ref{fig:stdy}) shows the {\it Airy peak} for three
velocities: $v = 0$ (static approximation), $v = 1 \lambda / s$,
and $v = 2.6 \lambda / s$, with ${\cal F} = 4000$. The curve
labeled {\it static}, corresponding to $v = 0$, was generated by
neglecting the travel time of light, i.e the cavity has reached 
its equilibrium point at each step. The two other curves, 
on the other hand, were simulated
according to the dynamical model here presented. Notice how the
main peak height decreases, its width increases and its position
shifts ahead of the resonance. These changes are greater for
larger velocities.

\section{The speed measurement}

\begin{figure}
\centering
\epsfig{file=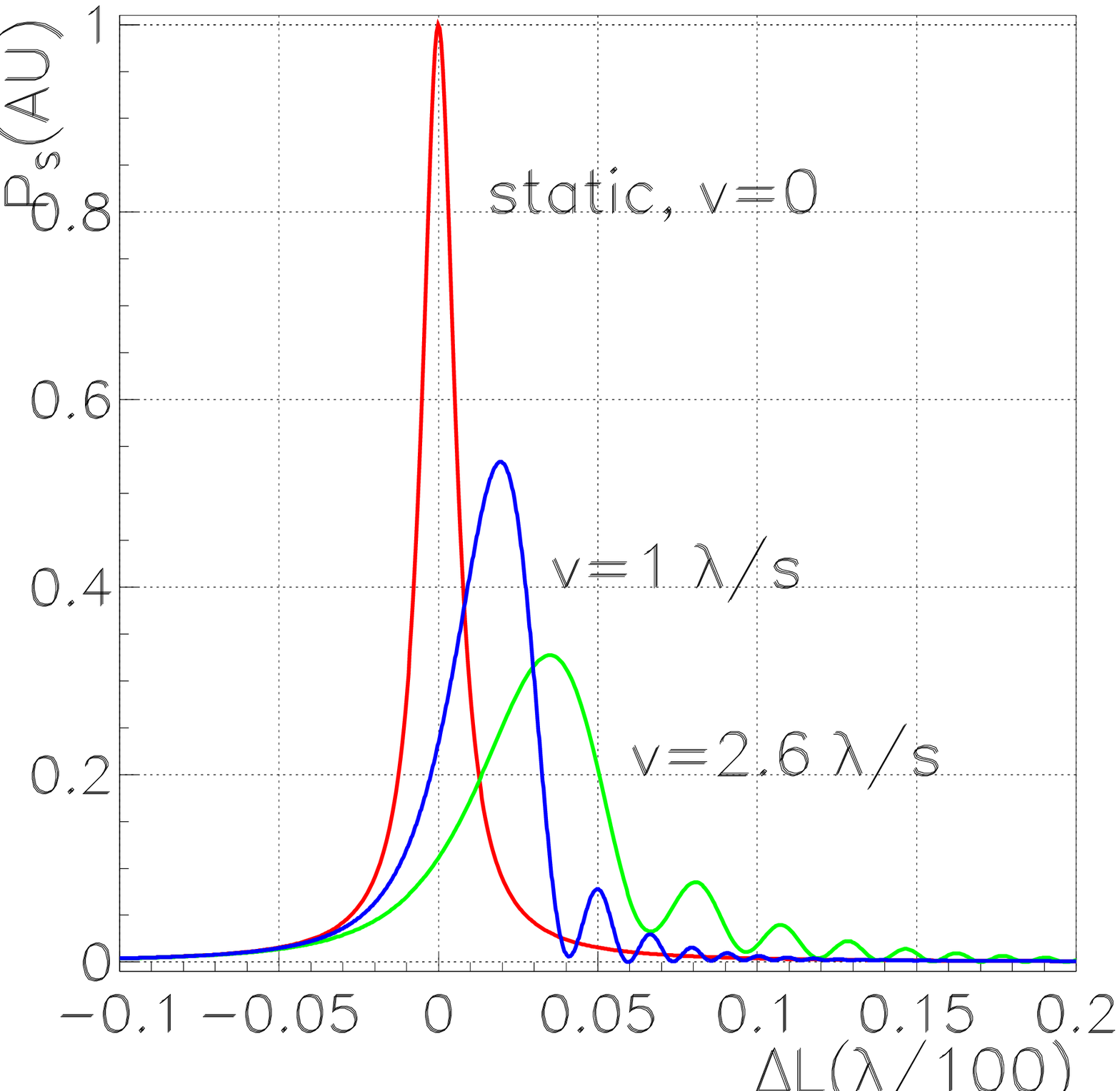,width=12cm}
\caption{Matone et al., Phys.Lett.A} \label{fig:stdy}
\end{figure}

\begin{figure}
\centering
\epsfig{file=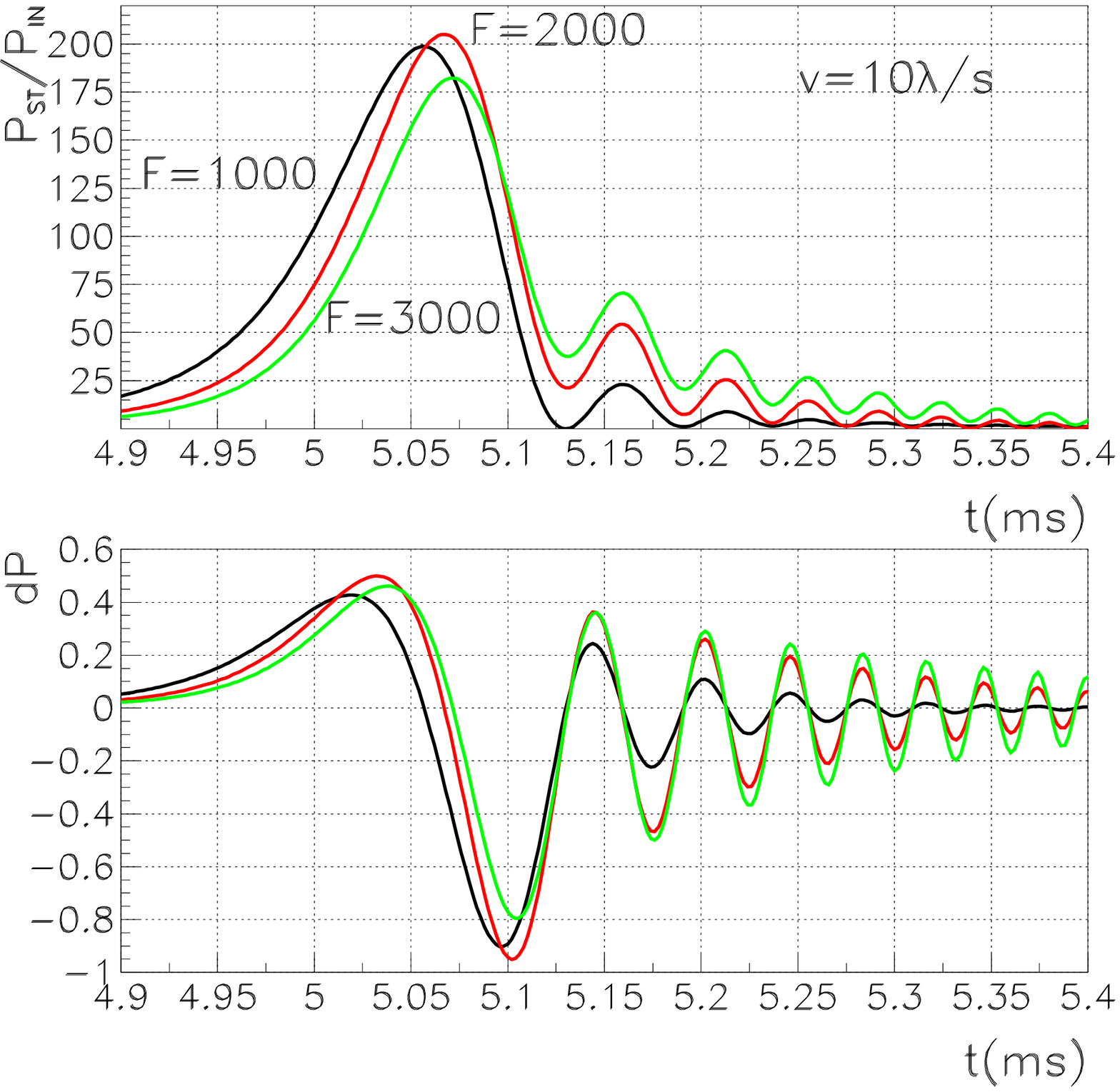,width=12cm}
\caption{Matone et al., Phys.Lett.A}
\label{fig:ex10a}
\end{figure}

We would now like to discuss a property of the ringing effect observed from
the simulation runs. Fig.(\ref{fig:stdy}) graphs the stored power
as a function of cavity length for a given finesse and for
different values of velocity. We can now plot the stored power as
a function of time, setting the velocity to a fixed value, but
varying the finesse. One example is given in
fig.(\ref{fig:ex10a}). The top graph of this figure shows the
stored power as a function of time, for an expansion rate set to
$v = 10 \lambda / s$, for three different finesse values: ${\cal
F} = 1000$, $2000$, and $3000$. The bottom
graph is the curves' time derivative. From these plots, we remark
a particular characteristic of the phenomenon: the position of the
minima and maxima, with the exception of the main peak, are
almost independent from the finesse value.

Furthermore, going back to fig.(\ref{fig:stdy}), we can now note
that the derivative zeros depend only on the relative mirror
velocity. 
The simulation output shown in fig.(\ref{fig:ex10a}) not only
shows how the derivative zeros are independent, at least to first
approximation, from the finesse, but it also shows a particular
regularity in the spacing between the minima and maxima.

\begin{figure}
\centering
\epsfig{file=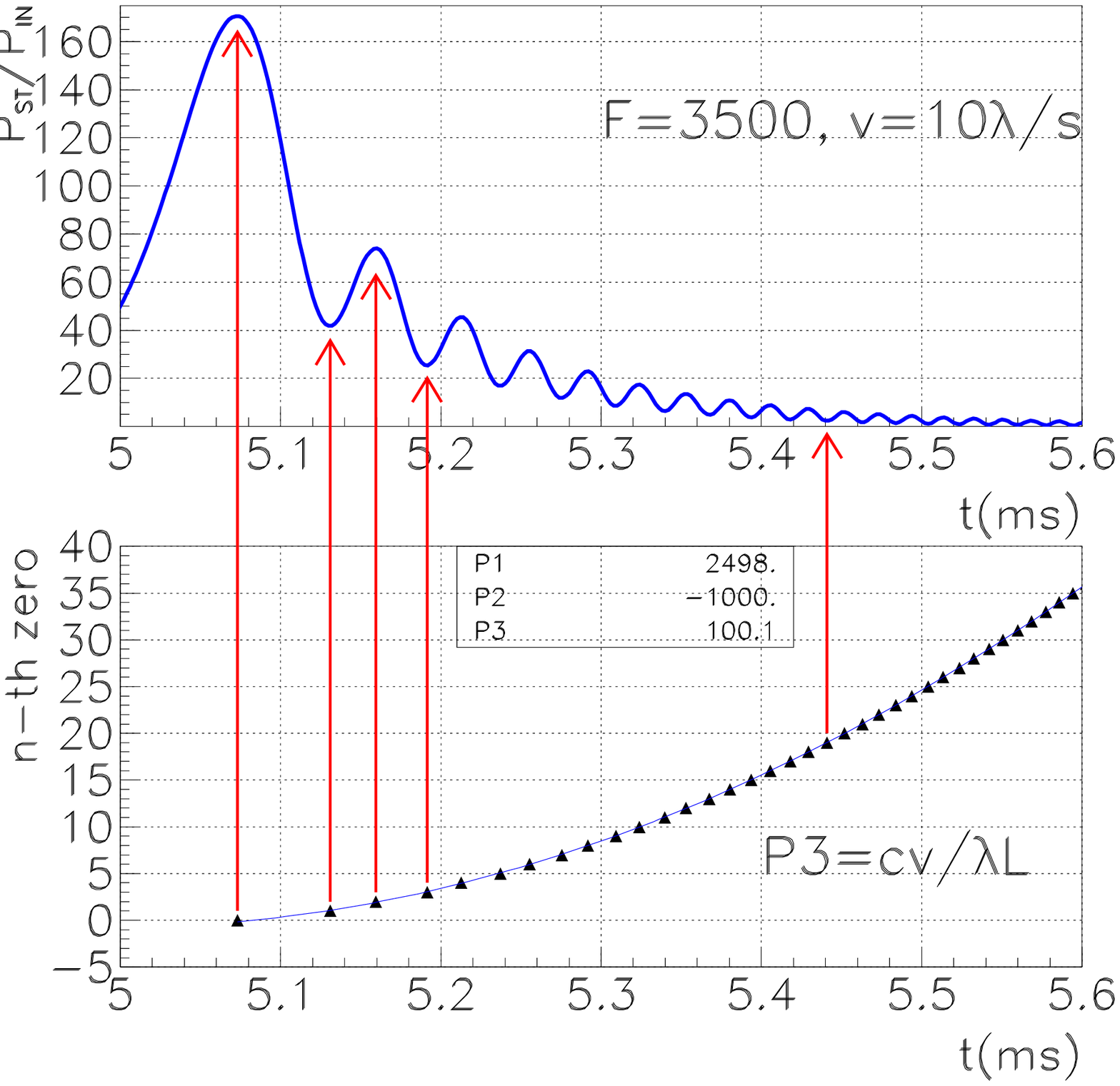,width=12cm}
\caption{Matone et al., Phys.Lett.A} \label{fig:ex10bis}
\end{figure}

The upper graph of fig.(\ref{fig:ex10bis}) shows the simulated
stored power of a cavity with ${\cal F} = 3500$, expanding at a
rate $10 \lambda/s$. Let the position of the curve's derivative
zeros, $t_{zero}$, be labeled by the index $n$, so that,
for the first zero, positioned at $t_{zero} \simeq 5.07 
ms$, $n = 0$, for the second zero, located at $t_{zero}
\simeq 5.103 ms$, $n = 1$ and so on. Then, the bottom graph of
fig.(\ref{fig:ex10bis}) shows the plot of index $n$ as a function
of time. We remark that the $n$-th zero of the derivative is a
quadratic function of the zero crossing time $t_{zero}$:
$n \propto t_{zero}^{2}$.
By fitting the simulation outputs to the expression
$ n_{zero} = p_1 + p_2 t_{zero} + p_3 t_{zero}^{2}$
where $p_{1,2,3}$ are fitting parameters, we empirically found that the
coefficient $p_3$ can be written as $ p_3 = c v / \lambda L$
where $L$ is the cavity length and $v$ is the cavity expansion
rate (an example is shown in the bottom graph of
fig.(\ref{fig:ex10bis})). Therefore, an estimate of
coefficient $p_3$ would also give
us an estimate of the relative velocity $v$.

\begin{figure}
\centering
\epsfig{file=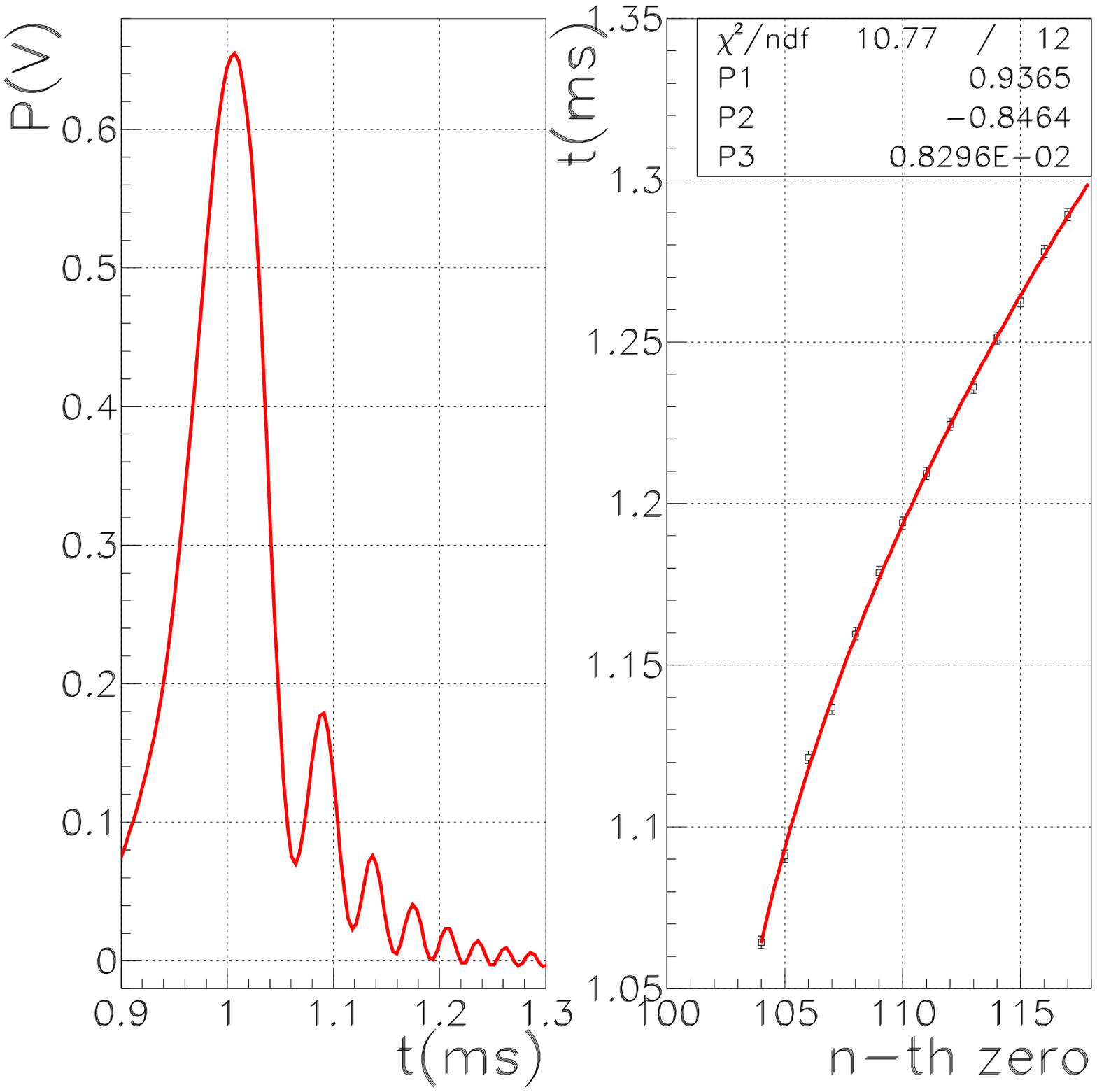,width=12cm}
\caption{Matone et al., Phys.Lett.A} \label{fig:tr123c}
\end{figure}

Fig.(\ref{fig:tr123c}) shows the results of a fit on a measured event. 
Notice how the parabolic behavior is in agreement with the
experimental points, resulting in a measured
speed of $12.8 \pm 10^{-2}\mu m/s$.

\section{The Finesse measurement}

\begin{figure}
\centering
\epsfig{file=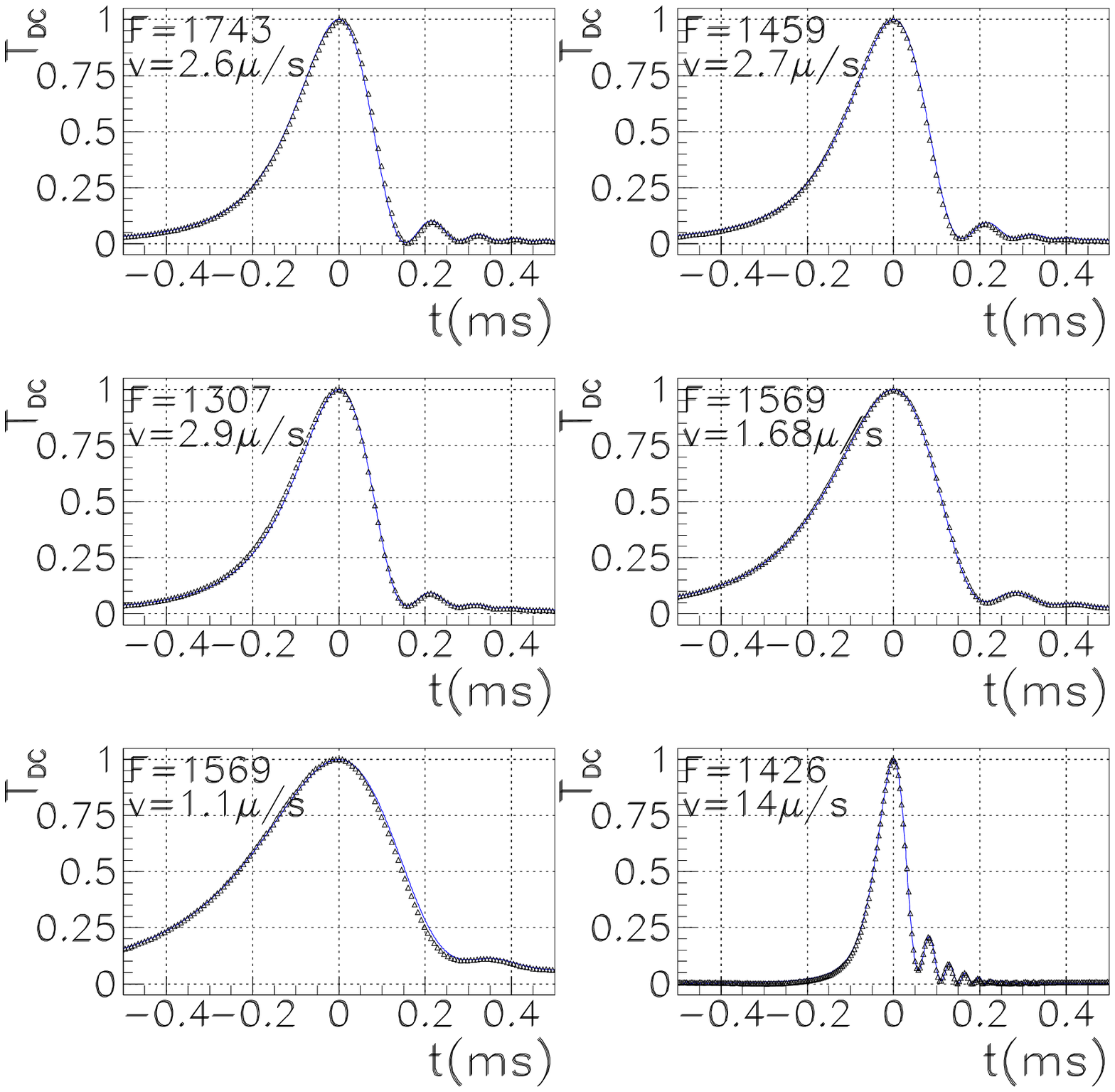,width=12cm}
\caption{Matone et al., Phys.Lett.A} \label{fig:fit_sie1}
\end{figure}

Once the speed is extracted, it is possible to fit the measurements
with the simulation's output to find the remaining parameter: the finesse.
The fit results of a set of measurements, six of which are shown
in fig.(\ref{fig:fit_sie1}), led to a mean finesse of
$ \overline{{\cal F}} = 1554 \pm 160 $, a value 
later confirmed by a measurement of the cavity pole.
The ten percent precision on the finesse measurement is most probably
due to the alignment state of the cavity, as suggested by simulation
studies.

\section{Conclusion}

An analysis of the optical ringing effect, observed on
the VIRGO mode-cleaner prototype in Orsay, was here presented.
We investigated a method to extract, from the oscillatory behavior,
both the relative mirror speed and the finesse of the system.

The numerical results showed how the position of
the oscillations' minima and maxima, when plotted as a function
of time, weakly depend on the finesse value and are completely 
determined by the cavity expansion rate as the resonance is
being crossed. In particular, we showed how a simple empirical 
formula can determine the cavity expansion rate by observing
these minima and maxima. 

Once the speed was reconstructed, it was possible to fit
the measurements with the simulation's output
and estimate the cavity's finesse to
$\overline{{\cal F}} = 1554 \pm 160$.

The present letter gives an alternative method to the finesse measurement
for a suspended cavity. The simplicity in the velocity reconstruction 
algorithm may be useful for a future control system 
capable of guiding the cavity into lock.


\newpage

{\bf Figure Captions}

Fig.1: The observed {\it ringing effect} on
the transmitted DC power of the MC30 prototype. The transmitted
power is shown as a function of time as the cavity length sweeps
the optical resonance at an unknown rate.\\

Fig.2: The model used for the study of the MC30 ringing effect.\\

Fig.3: The calculated Fabry-Perot transmitted power, with ${\cal F}
= 4000$, as a function of cavity length $\Delta L$ as the
resonance is swept at $v=0$ (static approximation), $v = 1 \lambda
/ s$, and $v = 2.6 \lambda / s$. In the figure, $\Delta L = 0$
corresponds to resonance.\\

Fig.4: The calculated stored power as
a function of time, with a fixed expansion rate set to $v = 10
\lambda / s$, for different finesse values: ${\cal F} = 1000$,
$2000$, and $3000$. Top graph: the stored
power. Bottom graph: the stored power time derivative.\\

Fig.5: The simulated stored power of a Fabry-Perot,
expanding at a constant rate $v = 10 \lambda/s$, with ${\cal F}
=3500$. Top graph: the stored power as a function of time. Bottom
graph: the index $n$, corresponding to the $n$-th derivative zero,
as a function of time. The curve is fit to the expression
$n = p_1 + p_2 t + p_3  t^2$. Notice that
$p_3  =  c v/\lambda L = 100 [1/ms^2]$.\\

Fig.6: Fit results for the mirror relative velocity
reconstruction.On the left: The measured DC transmitted power. On
the right: the plot of $t_{zero}$ as a function of index
$n$. The error bars correspond to half of the oscilloscope's
sampling time. The reconstructed speed is 
$12.8 \pm 10^{-2}\mu m/s$.\\

Fig.7: The observed ringing effect for the MC30 prototype:
measurements and fits.
The finesse and velocity values are shown for each graph.
$T_{DC}$ in arbitrary units.\\

\newpage

\newpage

\newpage

\newpage

\newpage

\newpage

\newpage

\begin{thebibliography}{9}
\bibitem{virgo} C. Bradaschia et al, {\it Nucl. Intrum. Meth. Phys. Res. A}
 {\bf 289,} 518 (1990).

\bibitem{ligo} A. Abramovici et al, {\it Science} {\bf 256,} 325 (1992).

\bibitem{geo} K. Danzmann et al, {\it Internal Report MPQ} {\bf 190} (1994).

\bibitem{tama} N.Kanda et al, {\it Proceedings of second Workshop of
Gravitational Wave Data Analysis}, Orsay 1997

\bibitem{mc} M.Barsuglia et al., {\it submitted to Rev.Sci.In}

\bibitem{lefloch}
J.P\'er\^ome et al., {\it J.Opt.Soc.Am.B}, {\bf 14} 2811 (1997)

\bibitem{stephane}
S. T'Jampens, {\it Rapport de Stage de Licence au LAL} (1996)


\end{thebibliography}
\end{document}